\documentclass[11pt]{article}
\usepackage{a4,graphicx,epsfig,amsfonts,amsthm}
\usepackage{subfigure}
\usepackage[rflt]{floatflt}
\usepackage{amsmath}
\usepackage{hyperref}
\usepackage{color}

\newcommand{\bkappa}{\mbox{\boldmath $\kappa$}}
\newcommand{\bk}{{\bf k}}

\newcommand{\bq}{{\bf q}}

\def\lsim{\mathrel{\rlap{\lower4pt\hbox{\hskip1pt$\sim$}}
    \raise1pt\hbox{$<$}}}         

\def\gsim{\mathrel{\rlap{\lower4pt\hbox{\hskip1pt$\sim$}}
    \raise1pt\hbox{$>$}}}         

\begin{document}

~\vskip 2cm\begin{center}
\bigskip
{\Large \bf \boldmath
Higgs Central Exclusive Production} \\
\bigskip
\bigskip
{\large J.R.~Cudell $^{a,}$\footnote{jr.cudell@ulg.ac.be}, A.~Dechambre $^{a,}$\footnote{alice.dechambre@ulg.ac.be}, O.~F.~Hern\'andez 
$^{b,}$\footnote{oscarh@physics.mcgill.ca; Permanent Address:
Marianopolis College, 4873 Westmount Ave., Westmount, QC H3Y 1X9
1W1 }}
\\
\vspace*{1cm}
{{\it $^{a}$ \small IFPA, D\'ep. AGO, Universit\'e de Li\`ege, Sart Tilman, 4000 Li\`ege, Belgium\\\
      $^{b}$ \small Physics Dept., McGill University, 3600 University St., Montr\'eal, Qu\'ebec,Canada, H3A2T8}}
\bigskip
\end{center}

\begin{center}
\bigskip (\today)
\vskip0.5cm {\Large Abstract\\} \vskip3truemm
\parbox[t]{\textwidth}{We tune the calculation of central exclusive Higgs production to the
recent CDF central exclusive dijet data, and predict the cross section for the exclusive production of Higgs boson at the LHC. It is always below 1~fb, and below 0.3~fb after experimental cuts.
}
\end{center}
\thispagestyle{empty}
\newpage
\setcounter{page}{1}

\section{Introduction}
The possibility of producing a Higgs boson in a diffractive event has been suggested a long time
ago \cite{Nachtmann,Landshoff}. The exclusive process was then recognized as a gold-plated signal
for a light Higgs boson \cite{Bj}, as it would show up irrespectively of its decay products. Many theorists 
built models predicting the rate of production, and because there is no factorisation theorem for 
production in a gap, the estimates varied wildly. Recently, CDF measured diffractive production
of jets in a mass range up to 130 GeV, so that one can now identify the ingredients needed to describe
the data.

The first attempt to embed Higgs boson production into a pomeron appeared in \cite{Landshoff}, which
used specific non-perturbative ingredients to model the pomeron. The first perturbative calculation
was performed in \cite{CH}, at the price of the introduction of an unknown proton impact factor.
Potentially large screening corrections were evaluated in \cite{Bj2}
in an eikonal formalism, and have been extended to any unitarisation scheme in \cite{Trosh,Frank}.
Finally, the necessity of a Sudakov form factor was noted in \cite{Collins}.

All these ingredients were incorporated in a series of papers by the KMR group~\cite{KMR}, who
used specific prescriptions for the Sudakov form factor, for the screening corrections, and for the
impact factor. In a previous paper~\cite{CDHI1} devoted to dijet exclusive production, we have shown
 that their choices are not the only ones possible, and we have proposed alternatives to several 
 of them:

\begin{figure}[h]
\centering \mbox {\subfigure[
]{\includegraphics[width=3.3cm]{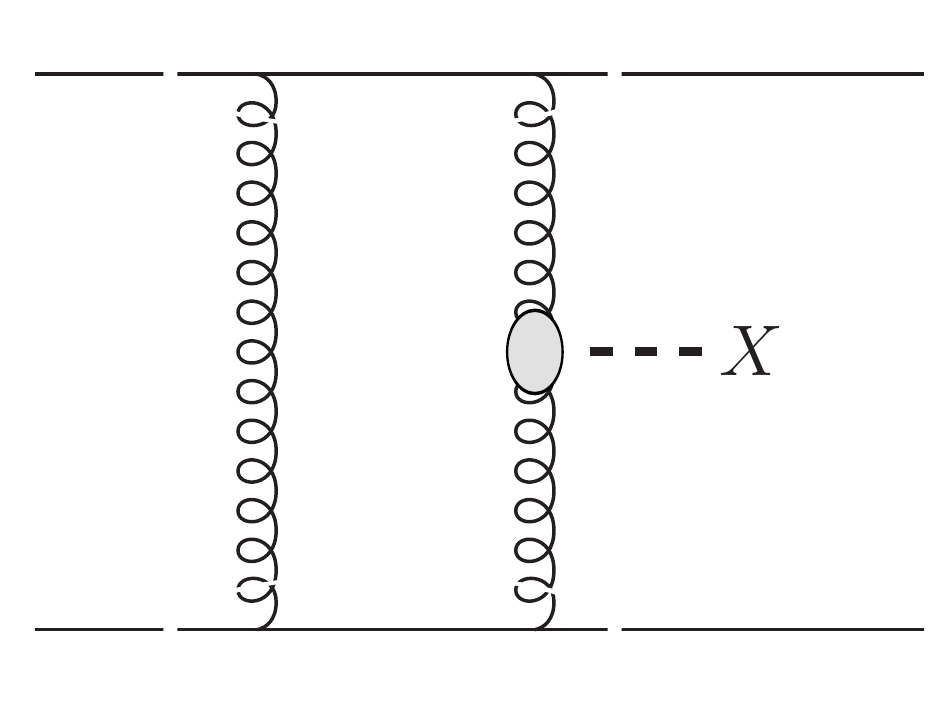}}
\quad \subfigure[]{\includegraphics[width=3.3cm]{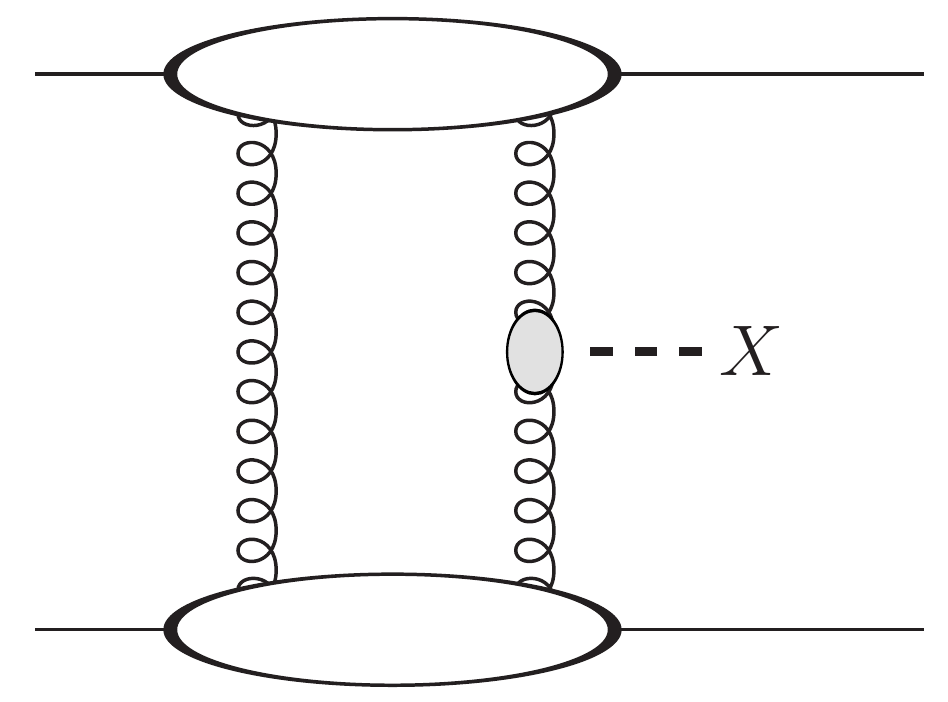}}\quad \subfigure[]{\includegraphics[width=3.3cm]{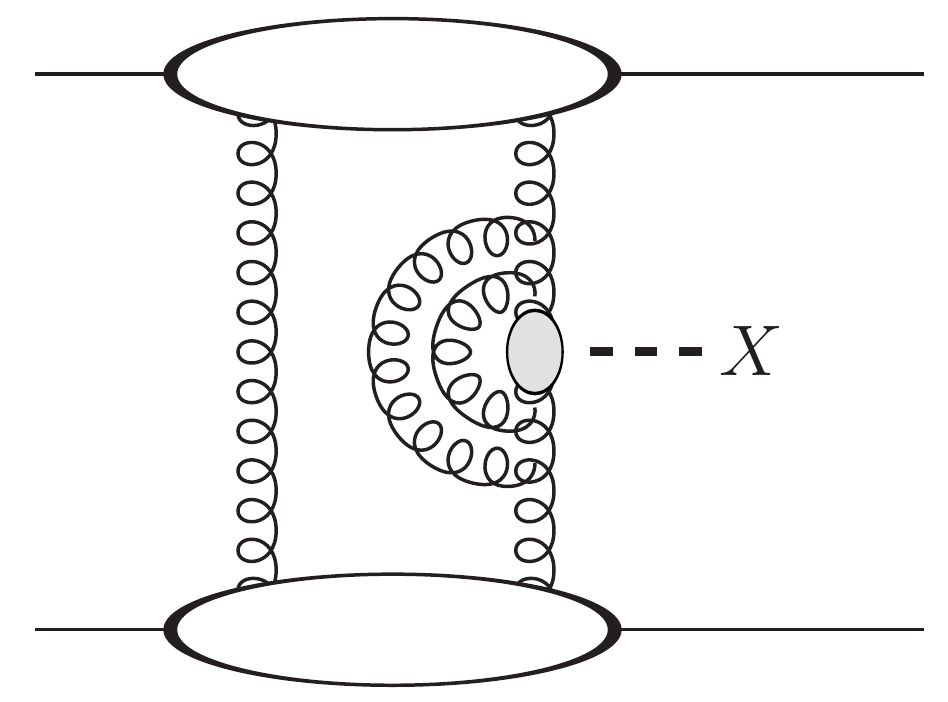}}\quad \subfigure[]{\includegraphics[width=3.3cm]{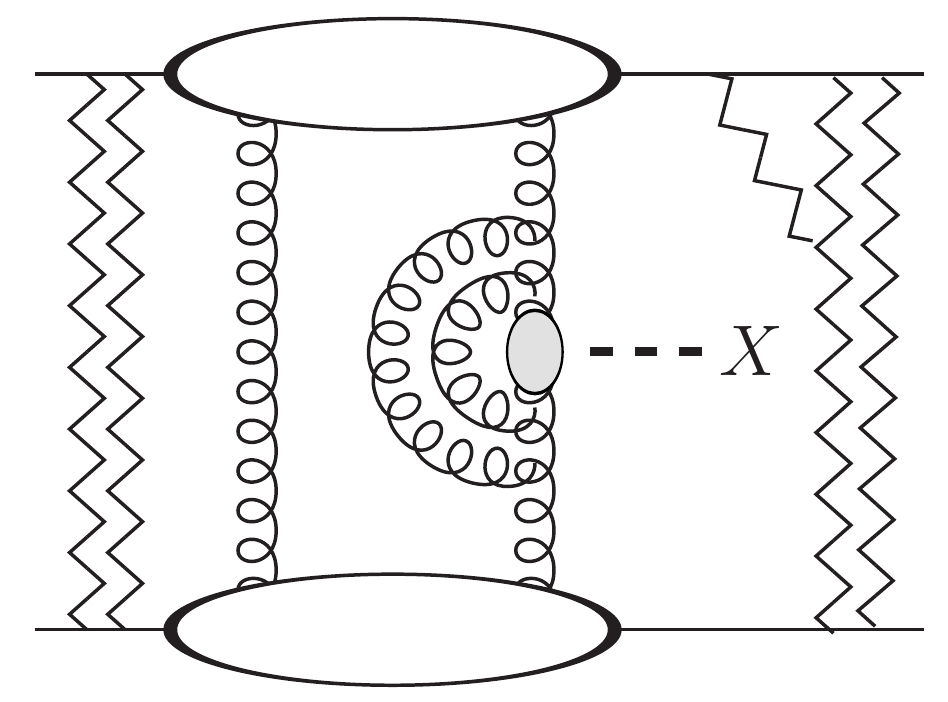}}}
\caption{Standard scheme of an exclusive cross section calculation with its various steps.}\label{fig_scheme}  
\end{figure}  

\begin{itemize}
\item Partonic amplitude: although the calculation lies partly in the infrared region, one usually 
calculates the amplitude in perturbative QCD by computing  the exchange of two gluons between two 
quarks, Fig.~\ref{fig_scheme}.(a). The second (screening) gluon is present in order to prevent a colour flow in the $t$~channel and allows the production of a colour singlet in the final state. In the present calculation the momentum lost by the proton is not 
neglected with respect to the loop momentum~$k$ as in~\cite{CDHI1}, and we have shown that this approximation overestimates the cross section by a factor~2. An exact transverse kinematics is thus kept.   
\item Impact factor: the infrared divergences of the QCD amplitude disappear because the protons is colour-neutral. This introduces  
 a proton impact factor, as illustrated in Fig.~\ref{fig_scheme}.(b). In the present paper,it is parametrised by an off-diagonal  skewed unintegrated gluon 
 density~\cite{CDHI1,Ivanov:2000cm} that allows for the exchange of soft gluons in the $t$ channel. 

\item Sudakov form factor: the virtual corrections, coming from the transition of gluons from low 
transverse momenta in the proton to hard ones in the Higgs-boson vertex, are accounted for by this factor, as shown in Fig.~\ref{fig_scheme}.c. The vertex corrections give double and single logarithms which must be resummed and the main uncertainty in the calculation comes from the constant terms that cannot be resummed although they can be large.
\item Gap survival probability: additional pomeron exchanges between the initial and final protons can occur and this leads to screening corrections~\cite{Bj2}, as shown in Fig.~\ref{fig_scheme}.(d)
\end{itemize}
We also showed that the calculation still largely lies in the non-perturbative region, so that more 
exotic choices are also possible~\cite{Bzdak}. The uncertainties coming from these various ingredients were studied, and in the following we shall use the CDF dijet results~\cite{CDHI1} to reduce the theoretical uncertainties on the 
Higgs-boson exclusive cross section. 

The first section of this letter summarizes the calculation of the Higgs-boson
exclusive cross section in the light of the dijet exclusive cross section. The second section 
presents the results using our new knowledge of the different ingredients.
 
\section{The Higgs-boson exclusive cross section}
\subsection{Kinematics and partonic amplitude}
\begin{figure}[h]
\begin{center}
\includegraphics[height=4cm]{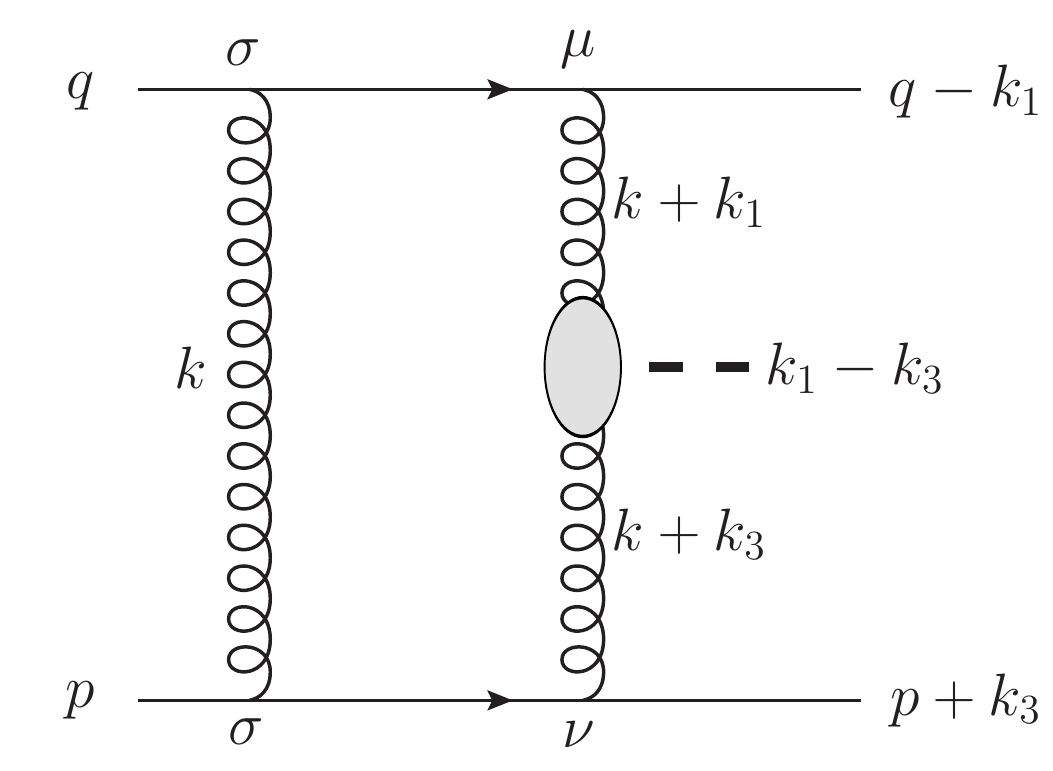} \caption{Kinematical conventions for the Higgs exclusive production at the partons level.}\label{fig_kinematic}
\end{center}
\end{figure}

The basis of the calculation is the partonic subprocess $qq\to q+H+q$, and involves the evaluation of the diagram 
shown in Fig.~\ref{fig_kinematic} and the corresponding cross diagram as well as the diagrams in which the Higgs is emitted from the other gluon.
We neglect the quark masses, their initial transverse momenta and write their four-momenta 
as~$q^{\mu}$ and~$p^{\mu}$ with~$s=2p\cdot q$. The total momentum transfers to the initial 
quarks are $-k_1$ and $k_3$ respectively and the momenta are decomposed onto the 
incoming quark directions using a Sudakov decomposition:
\begin{equation}
\begin{split}
k^{\mu}&=\alpha p^{\mu}+\beta q^{\mu}+\bk^{\mu},\\
k_1^{\mu}&=\alpha_1 p^{\mu}+\beta_1 q^{\mu}+\bk_1^{\mu},\\
k_3^{\mu}&=\alpha_3 p^{\mu}+\beta_3 q^{\mu}+\bk_3^{\mu},\\
\end{split}
\end{equation}
where all transverse vectors are denoted by bold letters.  The produced Higgs boson momentum is
\begin{equation}
k_H^{\mu}=(\alpha_1-\alpha_3)p^{\mu}+(\beta_1-\beta_3) q^{\mu}+(\bk_1-\bk_3)^{\mu}.\label{eq_higgsmomentum}
\end{equation}
The kinematical region of interest is multi-Regge kinematics where longitudinal momentum losses
are small and obey
\begin{equation}
1 \gg \alpha_i,\beta_i \gg \frac{|\bk_i^2|}{s}, \quad i=1,3;\quad \alpha,\beta\sim
\frac{|\bk^2|}{s}.
\end{equation} 
Let us also define the momenta of the colliding gluons
\begin{equation}
\kappa^{\mu}_1\equiv (k+k_1)^{\mu},\quad \kappa^{\mu}_3\equiv (k+k_3)^{\mu}.\\
\end{equation}

\begin{figure}[h]
\begin{center}
\includegraphics[height=1.6cm]{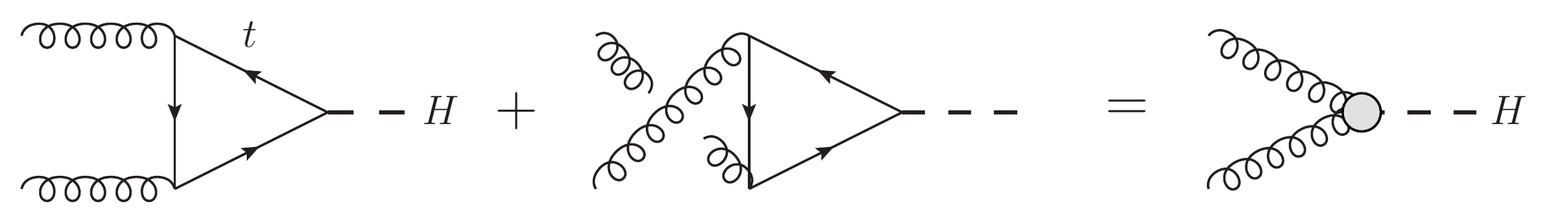} \caption{The Higgs-gluon-gluon vertex. Because the coupling of the Higgs to quarks is proportional to the quark mass, the top quark contribution dominates.}\label{fig_higgsvertex}
\end{center}
\end{figure}

At lowest order, the amplitude is purely imaginary and can be calculated using cutting rules. 
The parton-level amplitude then becomes\footnote{Note that ref.~\cite{CH} is missing a factor 2 
in the expression of the amplitude.}, in the high-$s$ limit~\cite{CH}:
\begin{equation}
\mathcal{M}_{qq}=g_s^4\frac{C}{2(2\pi)^2s}\int \frac{\mathrm{d}^2\bk}{\bkappa_1^2\bkappa_3^2\bk^2} (4 q^\sigma q_\mu)V^{\mu\nu} (4 p_\sigma p_\nu),\label{Mqq}
\end{equation}
where $g_s$ is the coupling constant between quarks and gluons, which will be 
later re-absorbed in the impact factor, $C$=2/9 is the colour factor for the two-gluons exchange and $V^{\mu\nu}$ is the Higgs-gluon-gluon effective vertex. The effective coupling between gluons and the Higgs boson can be obtained by calculating the diagrams in Fig.~\ref{fig_higgsvertex}.
This was done in~\cite{CH} where it was shown that the vertex can be written
\begin{equation}
 V^{\mu\nu}=\delta^{ab}(\kappa_1.\kappa_3\,g^{\mu\nu}-\kappa_1^\nu\kappa_3^{\mu})\frac{W_1}{m^2_H}+(\kappa_1^2\kappa_3^2\,g^{\mu\nu}+\kappa_1^{\mu}\kappa_3^{\nu}\,\kappa_1.\kappa_3-\kappa_1^{\mu}\kappa_1^{\nu}\,\kappa^2_3-\kappa_3^{\mu}\kappa_3^{\nu}\,\kappa_1^2)\frac{W_2}{m^4_H},
\end{equation}
where 
\begin{equation}
W_i=\frac{(\sqrt{2}G_f)^{1/2}g^2m^2_H}{12\pi^2}N_i.
\end{equation}
with $g$ the hard coupling constant, $G_f$ the Fermi constant and $m_h$ the Higgs mass.
The expression of $N_1$ and $N_2$ come from the integration over the quark loop in the particular 
kinematics of production in a rapidity gap $\bkappa_1^2,\bkappa_3^2\ll m^2_H$. Defining $a\equiv 
m_H^2/m_t^2$, one obtains
\begin{equation}
N_1=\frac{6}{a}\left[1+\left(1-\frac{4}{a}\right)\mathrm{arctan}^2\left(\sqrt{\frac{a}{4-
a}}\right)\right].
\end{equation}
The second part of the vertex, that depends of~$N_2$, can be neglected because the tensor 
structure will always give contributions proportional to~$\kappa_1^2$ or~$\kappa_3^2$ that are small
compared to the factors $s$ generated by the first tensor structure, and because $|N_2|^2$ is always
 less than~30$\%$ of~$|N_1|^2$  for Higgs masses below 1~TeV. 

Considering now the phase space available for the final Higgs boson and the above expression for the vertex, 
one can write the partonic differential cross section in the large-$s$ limit as
\begin{equation}\label{eq_partonlevel}
\mathrm{d}\sigma_{qq}=\frac{2C^2g^8}{(2\pi)^9}\,s\,\frac{\mathrm{d}\beta_1\mathrm{d}\alpha_3}
{(1-\beta_1)(1+\alpha_3)}\mathrm{d}^2\bk_1\mathrm{d}^2\bk_3
\left|\int\frac{\mathrm{d}^2\bk}
{\bk^2\bkappa_1^2\bkappa_3^2}
(\bkappa_1\cdot\bkappa_3)\frac{W_1}{m_H^2}\right|^2 \delta(k^2_H-m^2_H).
\end{equation}

The final delta function is used to put the Higgs on shell and to perform the integral over~$\alpha_3$. One can show using~Eq.~(\ref{eq_higgsmomentum}) and the on-shell condition on the final 
quarks that
\begin{equation}
\alpha_3\sim -\frac{m_H^2}{s\beta_1}.
\end{equation}

The above expression, Eq.~(\ref{eq_partonlevel}),  is singular at the partonic level when one of the exchanged gluons goes on-shell,~{\it i.e} if~$\bk^2\bkappa^2_1\bkappa^2_3\to$~0. This divergence disappears when one takes into account the fact that protons are colour singlets. This introduces 
an impact factor which furthermore describes the response of the proton to a transfer of momentum.
 
\subsection{Soft and Higher-Order Corrections}
The lowest-order partonic amplitude is the only piece of the calculation that can be exactly
calculated. The higher-order corrections (Sudakov form factor) and the soft corrections (impact factors 
and screening) can only be estimated. One can also use other exclusive data to constrain
the theory, and the only other input in a similar kinematical range is the dijet cross section from
the TeVatron. Hence we shall use our previous study \cite{CDHI1} to constrain the calculation.

\subsubsection{Impact Factor}
The wave-function overlap between the initial-state and final-state protons, after an exchange of $t$-channel
gluons of transverse momenta $\bk_a$ and $\bk_b$ and of longitudinal momentum fractions $x_a$ and $x_b$, 
leads to an impact factor $\Phi(x_a,x_b;\bk_a,\bk_b)$ that has the 
properties \cite{SoperGunion}
\begin{equation}
\Phi(x_a,x_b;0,\bk_b)=\Phi(x_a,x_b;\bk_a,0)=0.
\label{eq_impactfactor}
\end{equation}
The amplitude (\ref{Mqq}) for Higgs-boson production becomes for proton-proton scattering 
\begin{equation}
\mathcal{M}_{pp}=g^4\frac{C}{2(2\pi)^2s}\int \frac{\mathrm{d}^2\bk}{\bkappa_1^2\bkappa_3^2\bk^2} 
(4 q^\sigma q_\mu)V^{\mu\nu} (4 p_\sigma p_\nu)
 \Phi(\beta,\beta_1;\bk,\bkappa_1) \Phi(\alpha,\alpha_3;\bk,\bkappa_3)
 \label{mpp}
\end{equation}

To this, one must add an $s$-dependent factor that comes from the gluon content of
the proton (or equivalently from the pomeron trajectory), so that $\Phi$ is a function of
the longitudinal momentum fractions of the gluons. As the spectator gluon of momentum~$k$ has
a smaller momentum fraction than that of the gluon linked to the Higgs vertex, we assume that
the latter momentum fraction is dominant, and take $\alpha=\beta=0$ in Eq.~(\ref{mpp}).

Although the property~(\ref{eq_impactfactor}) is true in general, the precise form of $\Phi$ is
unknown. In \cite{CDHI1}, we tested several possibilities, and found that the shape of the $E_T$ 
distribution of the CDF dijet data was best reproduced if we assumed an $x$-dependence similar to that coming
from unintegrated gluon distributions. Hence we built the impact factor using three ingredients:
\begin{itemize}
\item The diagonal unintegrated gluon distributions $\mathcal{F}(x,Q^2)$ is built in \cite{Ivanov:2000cm}. 
They are 
built from a hard part, based on differentiation of conventional gluon densities 
(GRV~\cite{Gluck:1998xa}, MRS~\cite{Martin:1998np} or CTEQ~\cite{Lai:1997vu} LO fits), and from
a soft part, based on soft colour-singlet exchange inspired by the dipole form factor and allowing 
partons to enter the IR region. Four variants of the soft part were used, and their parameters were 
fixed by fitting the proton structure function 
$F_{2p}$ to the HERA data in a region of virtuality 0~$<$~$Q^2$~$<$~35~GeV$^2$ and for $x$ 
smaller 
than~10$^{-2}$. They are respectively named FIT-1, FIT-2, FIT-3 and FIT-4~\cite{Ivanov:2003iy,these}.
\item The skewed unintegrated distributions are obtained from the above by assuming that skewness, {\it 
i.e.} the non-zero longitudinal momentum transfer, can be taken into account
by using an effective longitudinal momentum fraction $x_g = 
0.41\,x_i$~\cite{Ivanov:2003iy,Shuvaev:1999ce}. 
\item The fully off-diagonal impact factors are obtained by multiplying the above
by a factor reproducing Eq.~(\ref{eq_impactfactor}) and going to~1 in the diagonal case, and by an exponential factor reproducing the proton~$t$ slope and shrinkage. One also chooses
the scale $Q$ to be the average of the incoming transverse momenta $Q^2=(\frac{\bk_a+\bk_b}{2})^2$. 
\end{itemize}
One then obtains for the upper proton:
\begin{equation}
\Phi(0,\beta_1;\bk,\bkappa_1)=\mathcal{F}\left(0.41\beta_1,(\bk+\frac{1}
{2}\bk_1)^2\right) 
\frac{2\bk^2\bkappa_1^2}{\bk^4+\bkappa^4_1}\quad e^{-\frac{1}{2}\left[B_0+2\alpha'\log\left(\frac{x_0}
{\beta_1}\right)\bk_1^2\right]},
\end{equation} 
with $B_0$=4~GeV$^{-2}$, $\alpha'$=0.25~GeV$^{-2}$ and $x_0$= 3.4~10$^{-4}$~\cite{CDHI1}.\\
These parametrisations provide a very good description of the charm contribution to the proton structure function~$F_{2p}$, of the longitudinal structure function $F_L$ and of diffractive vector meson 
production~\cite{Ivanov:2003iy,Ivanov:2004ax} as well as of the dijet exclusive cross section at the 
TeVatron~\cite{CDHI1}.  

\subsubsection{The Sudakov Form Factor}
During the exchange of gluons producing the Higgs boson, the gauge field goes from a long-distance 
configuration to a short-distance one. This is the situation in which, for running $\alpha_S$, large
logarithms arise from virtual diagrams such as those of Fig.~\ref{fig_scheme}.(c).
These logarithms become infinite if the external legs go on-shell, and it
is known that these infinities are canceled by the bremstrahlung diagrams, which are easier
to calculate. Hence, in order to evaluate the vertex corrections to $g^* g^*\to H$, one can
use the standard formula~\cite{livrerose}, assuming a point-like $ggH$ vertex:
\begin{equation}
T(\mu^2,\ell^2)=e^{-S(\mu^2,\ell^2)}, \quad S(\mu^2,\ell^2)=\int^{\mu^2}_{\ell^2}\frac{\mathrm{d}\bq^2}
{\bq^2}\frac{\alpha_s(\bq^2)}{2\pi}\int^{1-\Delta}_0\mathrm{d}z[zP_{gg}+N_fP_{gq}],
\end{equation}
where $\Delta=|\bq|/\mu$ and $P_{gg}$ and $P_{gq}$ are the splitting functions
\begin{equation}
P_{gg}(z)=2N_c\left[\frac{z}{1-z}+\frac{1-z}{z}+z(1-z)\right],\quad P_{gq}(z)=\frac{1}{2}\left[z^2+(1-
z)^2\right].
\end{equation}
The amplitude becomes
\begin{equation}
\mathcal{M}_{pp}=g^4\frac{C}{\pi^2}\int \mathrm{d}^2\bk \frac{\Phi(0,\beta_1,\bk,\bkappa_1) \Phi(0,|\alpha_3|,\bk,\bkappa_3)}{\bkappa_1^2\ \bkappa_3^2\ \bk^2} 
q_\mu V^{\mu\nu}p_\nu 
\sqrt{T(\mu^2,\bkappa_1^2)\ T(\mu^2,\bkappa_3^2)}.
\end{equation}
The Sudakov form factor is a loop correction to the vertex and includes an integral over the momentum in the loop,~{\it i.e.} over both the fraction of light-cone momentum~$z$ and the transverse momentum~$\bq^2$. The integration builds up double logarithms in the kinematical region where the virtual emission is soft and runs from a soft to a hard scale, respectively,~$\ell^2$ and $\mu^2$.  The upper scale is chosen according to the standard discussion of the Sudakov form factor where the hard point-like vertex is characterized by only one scale, the mass of the centrally produced system~$m_X$. This is the situation in the Higgs boson production for which the log structure of the Sudakov form factor can be calculated. It has been estimated
to single-log accuracy in~\cite{Kaidalov:2003ys} and recently in~\cite{Coughlin:2009tr}, the upper limit of the $z$-integration is then of the order of~$m_H$. The lower scale $\ell^2$ is proportional to the virtuality from which the evolution starts~$\bkappa_i$ and this leaves an uncertainty related to the constant terms which cannot be obtained from
bremstrahlung, and which are known not to exponentiate. For the purpose of this study, we perform the $z$ integration in the argument of the Sudakov correction~$S(\mu^2,\ell^2)$ and write its structure in three terms. After the second integration considering a running $\alpha_S$, the first will give the log contribution, the second the (log)(log) contribution and the third a constant term: 
\begin{equation}
S\propto \int^{\mu^2}_{\ell^2}\frac{\mathrm{d}\bq^2}
{\bq^2}\frac{\alpha_s(\bq^2)}{2\pi}\ \big[S_{\log} + S_{\log\log} + S_{const}\big]
\end{equation}
with
\begin{equation}
\begin{split}
S_{\log}&=-6\log{\Delta},\\
S_{\log\log}&=\frac{1}{3}N_f-\frac{11}{2},\\
S_{const}&=12\Delta-9\Delta^2+4\Delta^3-\frac{3}{2}\Delta^4+\frac{1}{2}N_f\Big[-\Delta+\Delta^2-\frac{1}{3}\Delta^3\Big].
\end{split}
\end{equation}
We have kept the latter contribution as in other exclusive models but we allow its variation by a factor~2 up or down. This constitutes the uncertainty coming from the constant terms in the Sudakov form factor. The effect of this change is an overall factor. \\


Note that perturbation theory suggests that the Sudakov form factor should be included 
in the definition
of the unintegrated gluon distribution. We do not do this here, because the use of an 
unintegrated 
gluon distribution is only an educated guess leading to the impact factor. There is no 
reason to
believe that the two objects are identical, as the exchanged spectator gluon cannot be 
assigned
to only one of the initial protons. 

\subsubsection{Gap-Survival Probability}
There is no factorisation theorem in diffractive exclusive production, so it may be 
important
to take initial- and final-state exchanges into account. This is at best tentative, as 
one
really needs to model the multiple-pomeron exchanges which account for the purely 
elastic cross
section, {\it i.e.} one needs to overcome the difficulty that lead the analytic S-matrix 
theory to a standstill.

Most of the present estimates are based on an eikonal scheme of unitarisation, following 
the
arguments presented by Bjorken a while ago~\cite{Bj}. They have been recently
generalised to an arbitrary unitarisation scheme~\cite{Frank}, for the case in which 
the production process happens at short distance. The gap survival probability can be
calculated as the $S$-matrix element, which factors in impact-parameter space and 
depends on the variables conjugate to $k_1$, $k_3$ and $k_1+k_3$. Such construct, and 
its transform back to momentum space, has never been implemented, and one usually treats 
the 
gap-survival probability as an overall factor. Several estimates can be found in the literature~\cite{Achilli:2007pn},  from 5$\%$ to 
15$\%$ at the 
TeVatron energy, and about a factor 2 lower at the LHC. These are conservative 
estimates, as the gap
survival at the LHC could be as large as 25 \% or smaller than 1 \%~\cite{Frank}. In the present study and in Sec.~\ref{section_results}, we assume a gap survival probability at the LHC two times smaller that the one used at the TeVatron as suggested by eikonal models~\cite{Achilli:2007pn}.

\subsection{First Estimation of the Higgs Boson Exclusive Cross Section}
Putting together all the above ingredients, we can estimate the Higgs boson cross 
section
for sets of reasonable theoretical parameters given in the second column of Table \ref{CDFparams}. 
\begin{table}
\begin{center}
\begin{tabular}{cccc}
\hline  
Physical Quantity  & Allowed Values & Fit to the &1 $\sigma$ Values\\
& & CDF data&\\
\hline
Parametrisation of the  &   \cite{Ivanov:2003iy,Ivanov:2004ax} & FIT-4  & FIT-4 \\
unintegrated gluon distribution& &&\\

Constant terms in Sudakov exponentiation  & C/2 to 2\,C &  C=1C & C/2 to 2\,C\\

Gap survival probability  &   4.5-18\%  & 10 \% & 10\% \\
\hline
\end{tabular}
\caption{Allowed range of parameters, and parameters reproducing the central values of the CDF exclusive dijet data. These will be used to predict the Higgs-boson exclusive cross section.}\label{CDFparams}
\end{center}
\end{table}
\begin{figure}[h]
\begin{center}
\includegraphics[height=7cm]{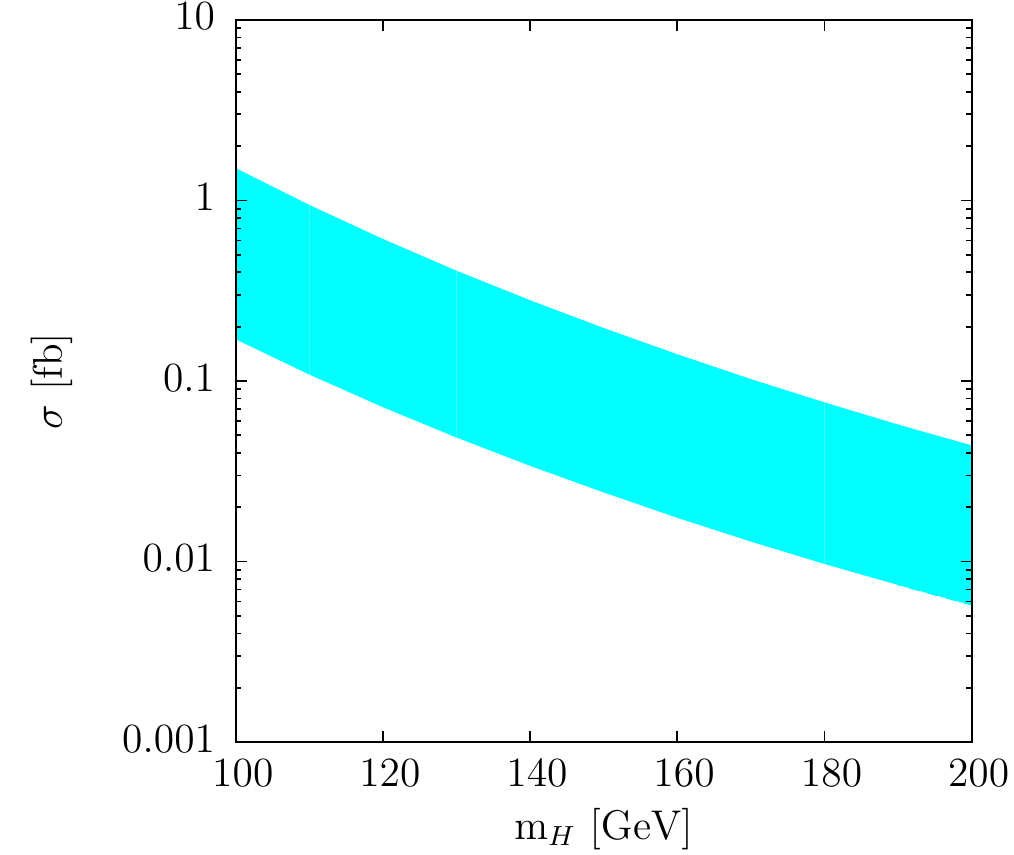} \caption{Prediction of the Higgs exclusive production cross section at 14 TeV without any experimental cuts and without tuning to CDF dijet data.}\label{fig_maxhiggs}
\end{center}
\end{figure}
and we see that
modest changes in the theoretical input can produce large differences in the predicted cross
section. The Higgs boson exclusive cross section in this model is below one femtobarns at
the LHC\footnote{Similarly small cross section has been obtained in references~\cite{Maciula:2010tv,Szczurek:2010jz}.}. The uncertainty is of one order of magnitude and the slope as a function of the Higgs mass is steeper than in the existing similar models~\cite{KMR,Maciula:2010tv}.

\section{From Dijet to Higgs}\label{section_jjtoH}
\begin{figure}[h]
\centering \mbox {\subfigure[
]{\includegraphics[width=3.6cm]{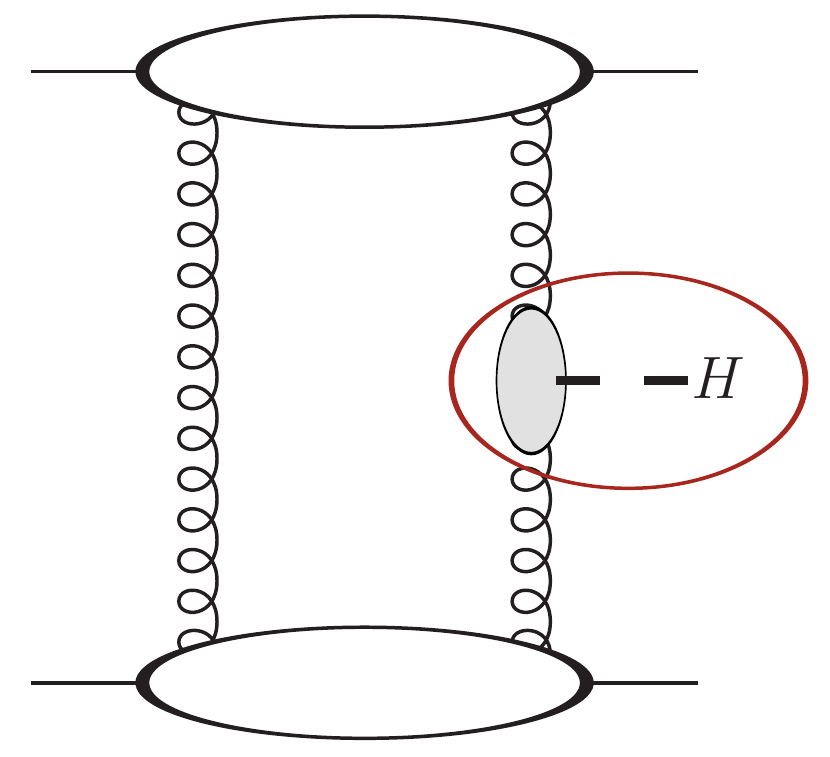}}
\quad\quad\quad\quad\quad\quad \subfigure[]{\includegraphics[width=3.6cm]{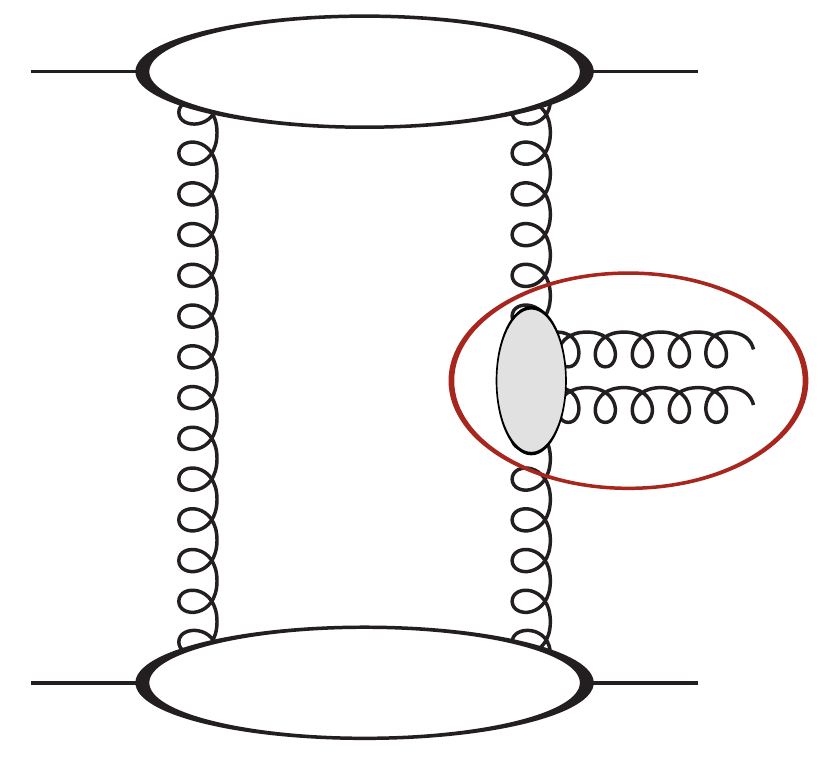}}}\caption{(a) Higgs exclusive production and (b) Dijet exclusive production}\label{fig_topology}  
\end{figure}

\begin{figure}[h]
\begin{center}
\includegraphics[height=3cm]{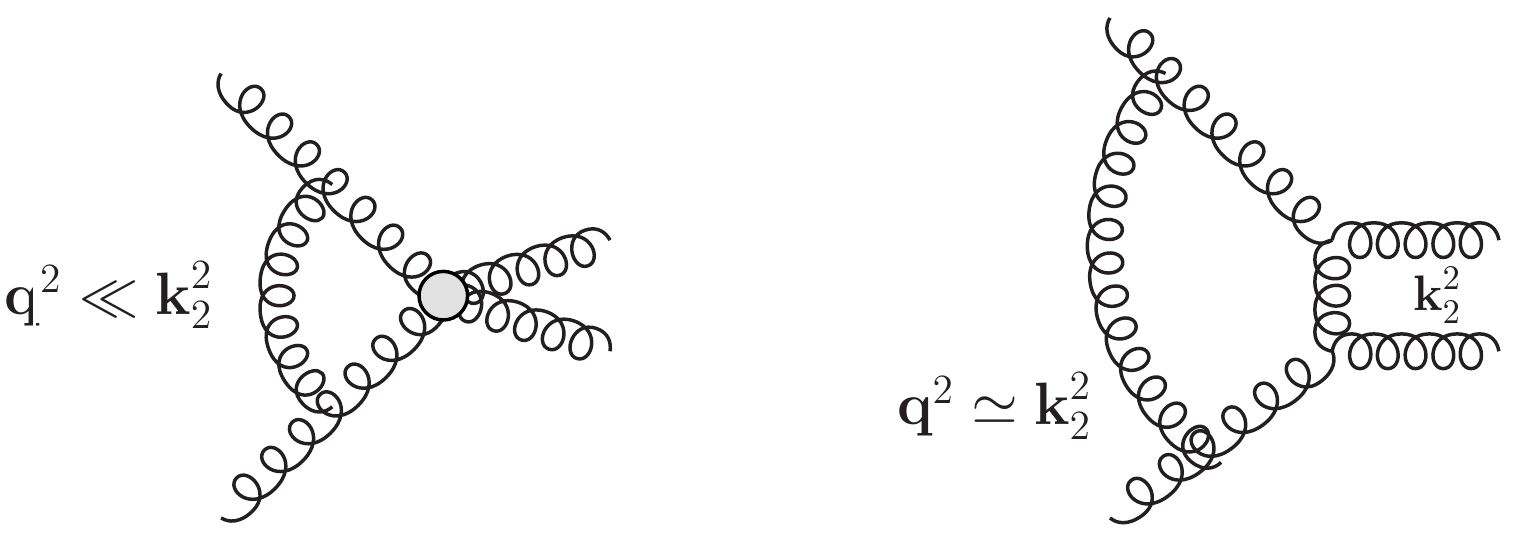} \caption{The two different kinematical regimes for colour-singlet dijet production. Left: The effective point-like vertex $\bq^2\ll\bk_2^2$. Right: The resolved vertex $\bq^2\simeq\bk_2^2$.}\label{fig_resovertex}
\end{center}
\end{figure}

One can in fact do a somewhat better job by constraining the model to reproduce existing data. The process of interest is the dijet exclusive production for which one of the diagrams is shown in Fig.~\ref{fig_topology}.b and for which the CDF collaboration published a measurement in 2008~\cite{Aaltonen:2007hs}.
These are important data for two reasons, firstly the calculations are expected to be almost identical, {\it i.e.} they differ only because of the hard vertices that produce the final particles. Indeed, diagrams in which the final-state gluons are produced by two different $t$-channel gluons can be safely neglected in the present kinematics as it contains one more hard propagator. In~\cite{IvanovDechambre}, we have evaluated its maximum contribution and have showed that it is safe to consider only those of Fig.~\ref{fig_topology}. Secondly, the dijet system reaches  masses of the order of 130~GeV, {\it i.e.} the region of mass where the Higgs boson is expected. The CDF data thus provide a good opportunity to tune the calculation of quasi-elastic production and to reduce the uncertainties that affect this family of process. The model of production of exclusive dijet is the same as the one presented in the first section and developed in~\cite{CDHI1}. However, the sources of uncertainties on both processes are different due to  the structure of the Sudakov form factor. One knows that a typical double-log-enhanced correction gets sizable logs only for a point-like vertex, or in other words, only if there is a large virtuality inside the effective vertex. This is the case in the Higgs boson production where the virtuality of the top quarks in the vertex shown in Fig.~\ref{fig_higgsvertex} is suffisuantly large in front of~$\bq^2$. On the contrary, in the dijet case, one can end up with the two situations pictured in Fig.~\ref{fig_resovertex}
In the left picture, the flow of transverse momentum $\bq^2\ll\bk_2^2$ in the loop does not change the amplitude while in the right one, the momentum~$\bq^2\simeq\bk_2^2$ is enough to resolve the vertex. The transverse logarithm does not build up leading to a complete change of the log structure of the amplitude. Consequently in the dijet exclusive production, the upper scale must be of the order of $\bk_2^2$ as it marks the limit beyond which the double-logs do not build up. This raises the question whether the dijet exclusive production data can be used to tune the Higgs boson calculation due to the extra uncertainty coming from the Sudakov form factor.

One can note that the gap survival probability is an overall factor that multiplies the cross section and a source of uncertainty similar to the one coming from the constant terms of the Sudakov form factor. Both uncertainties can be considered as a single one and cannot be reduced independently and this disavantage is reduced if one considers the ratio between the Higgs boson and the dijet exclusive cross section. It removes the uncertainty coming from the dependence of the cross section on the energy behaviour of the gap survival probability and underlines the importance of dijet data at the LHC. The uncertainty on the constant terms is reduced but not removed as the hard scale used in both processes is different ($\mu\sim E_T$ in the dijet case and $\mu\sim m_H$ in the Higgs case). This means that, whatever is the precision reached by the dijet data at the LHC, the precision on the theoretical prediction for the Higgs boson exclusive cross section will be smaller. \\

\begin{figure}[h]
\begin{center}
\includegraphics[height=8cm]{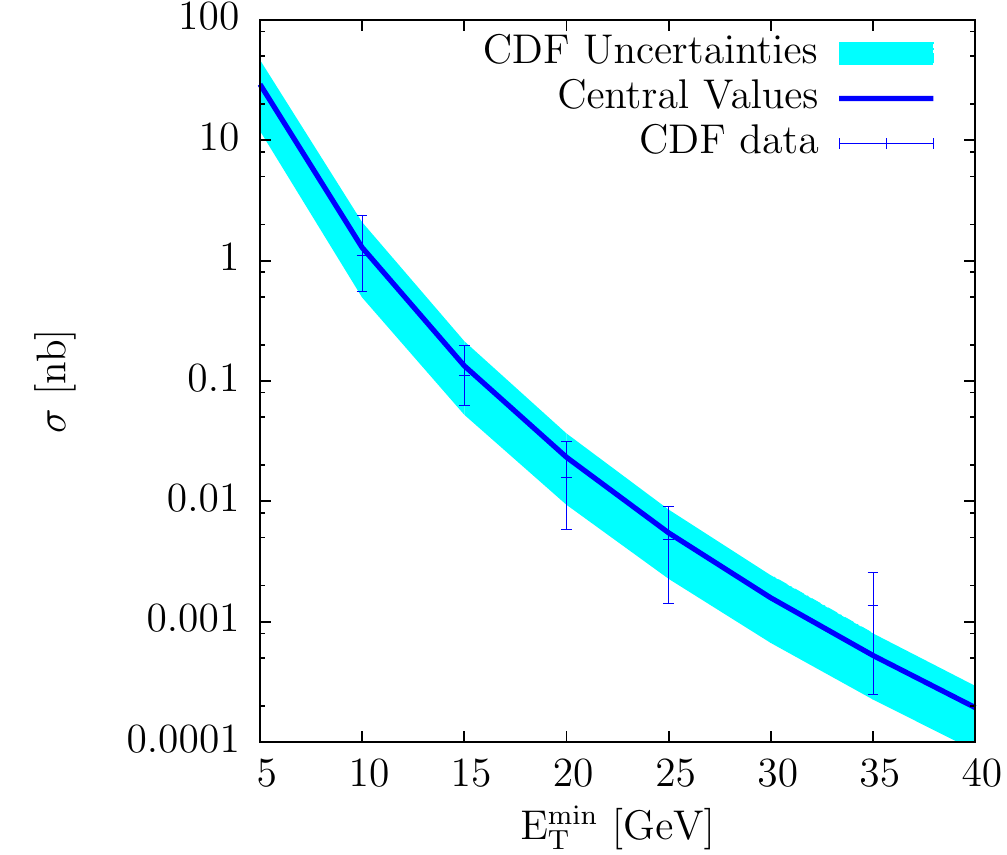} \caption{CDF Run II data on exclusive dijet, and a band of parametrisations that all go through the data. The curve corresponds to a  choice of parameters that best fit the central values and the band corresponds to parametrisations that go through the 1~$\sigma$ errors. Both include the effect of splash-out discussed in~\cite{CDHI1}.}\label{fig_CDFdata}
\end{center}
\end{figure}

The comparison between different parametrisations of the dijet exclusive production  model and the CDF data have led to a selection of parameters that reproduce the measurement well, as shown in Fig.~\ref{fig_CDFdata}. 
The central curve corresponds to one of the different choices of parameters that reproduce the central values of the data points and the band corresponds to parametrisations that are within the 1~$\sigma$ errors, also given in Table~\ref{CDFparams}. Using the knowledge of the TeVatron dijet data, we now keep the parametrisation of the unintegrated gluon distribution, the constant terms in the Sudakov exponentiation and the gap survival probability, we introduce them in the Higgs boson exclusive production calculation and present a prediction for the cross section using the parameters from the third and fourth columns of Table \ref{CDFparams}.

\par
\section{Final results}\label{section_results}
We now present the results for the cross section at the Higgs boson level but experimentally, what is detected in the central detector is the decay products of the Higgs, a~$b\bar{b}$ pair in the mass range considered. To obtain the cross section, the final jets have a priori to be reconstructed via a jet searching algorithm and this should be taken into account as the transverse energy of the parton is not equal to the transverse energy of the jet due to the energy lost when reconstructing the jet. The effects of the so-called splash-out were studied into details for gluon production in~\cite{these} but are not included here in the Higgs case. They are of course included in Fig.~\ref{fig_CDFdata}. Indeed, one can use the very forward detectors to reconstruct the mass of the central system from the measurement of the fraction of momentum lost by the initial hadrons. One in fact scans the mass domain to obtain the cross section without using the information coming from the central detector\footnote{Actually, one has to match the presence of hadrons in the forward detectors with an Higgs boson event in the central detector but the mass of the centrally-produced system and the cross section can be obtained using forward detectors only~\cite{Albrow:2008pn}. The effect of the Higgs decay and the reconstruction of the~$b\bar{b}$ final state were studied in the same model in reference~\cite{Dechambre:2011py}.}.\\ 

\begin{figure}[h]
\begin{center}
\includegraphics[height=4.3cm]{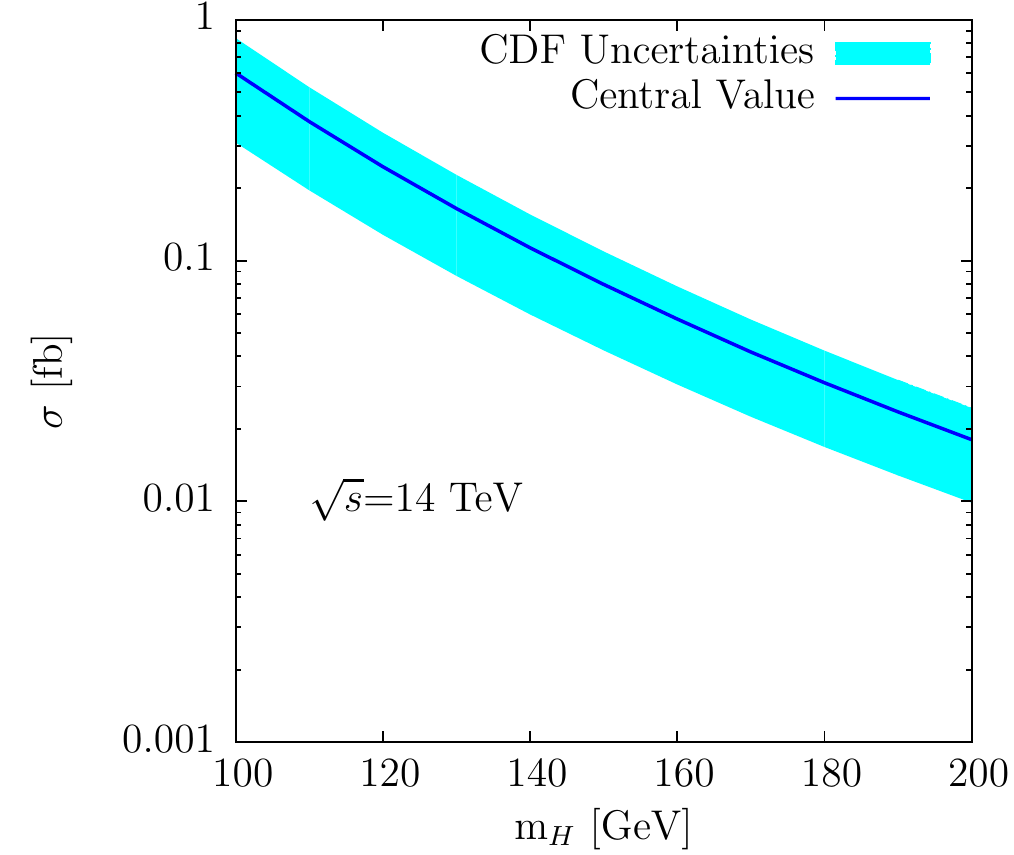} \includegraphics[height=4.3cm]{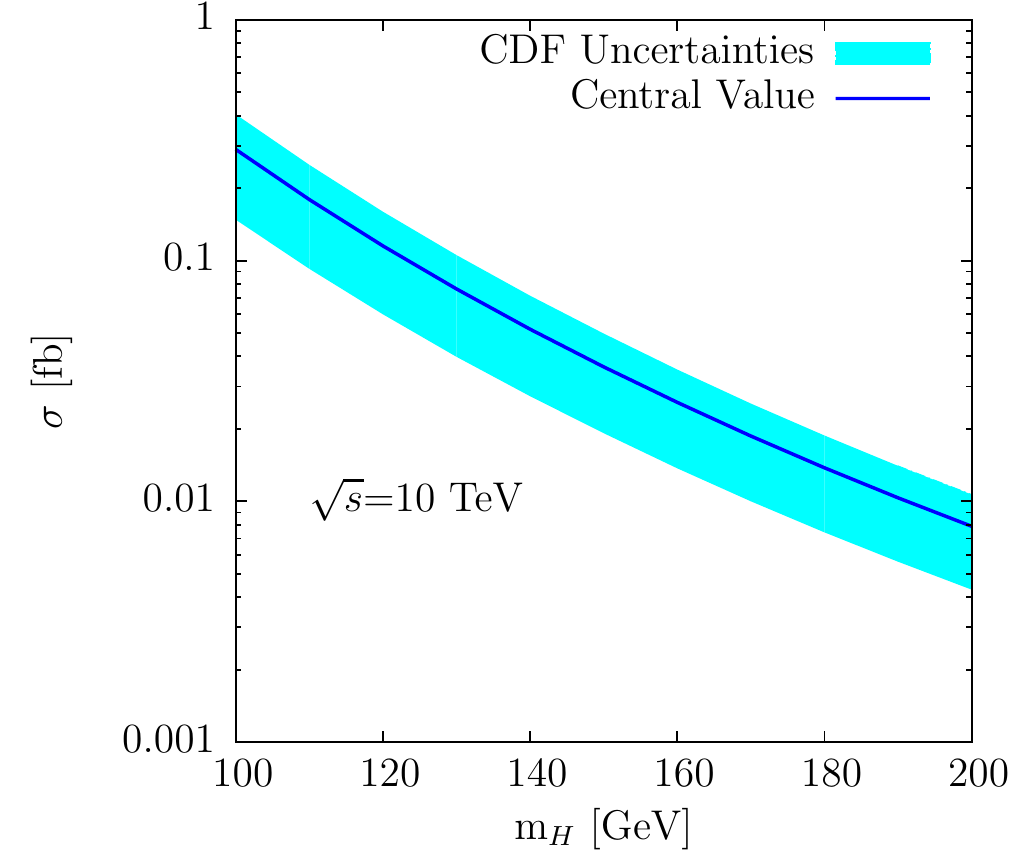} \includegraphics[height=4.3cm]{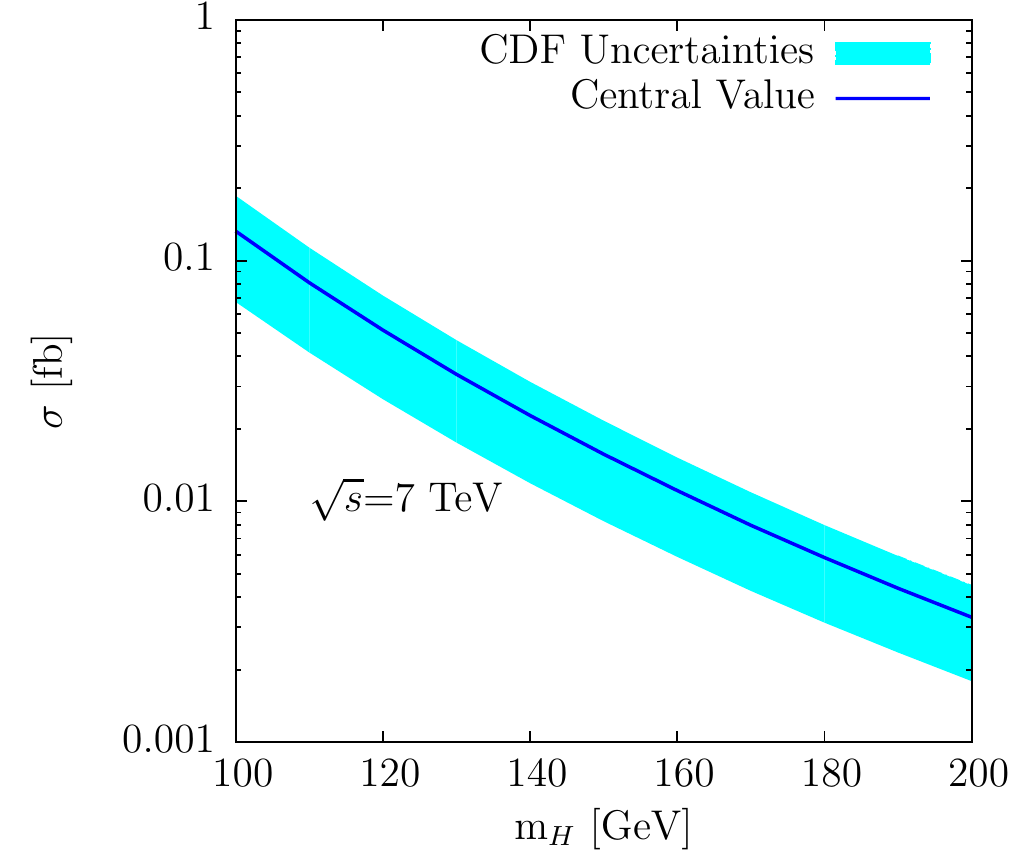} \\
\includegraphics[height=4.3cm]{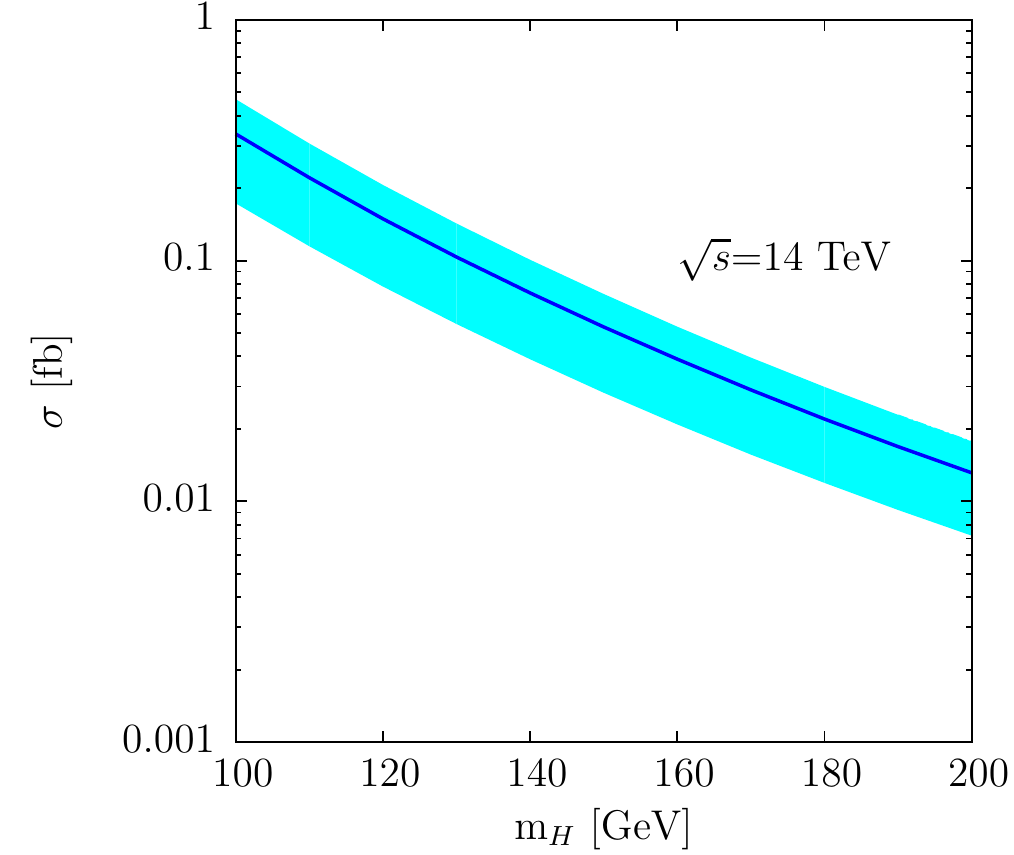} \includegraphics[height=4.3cm]{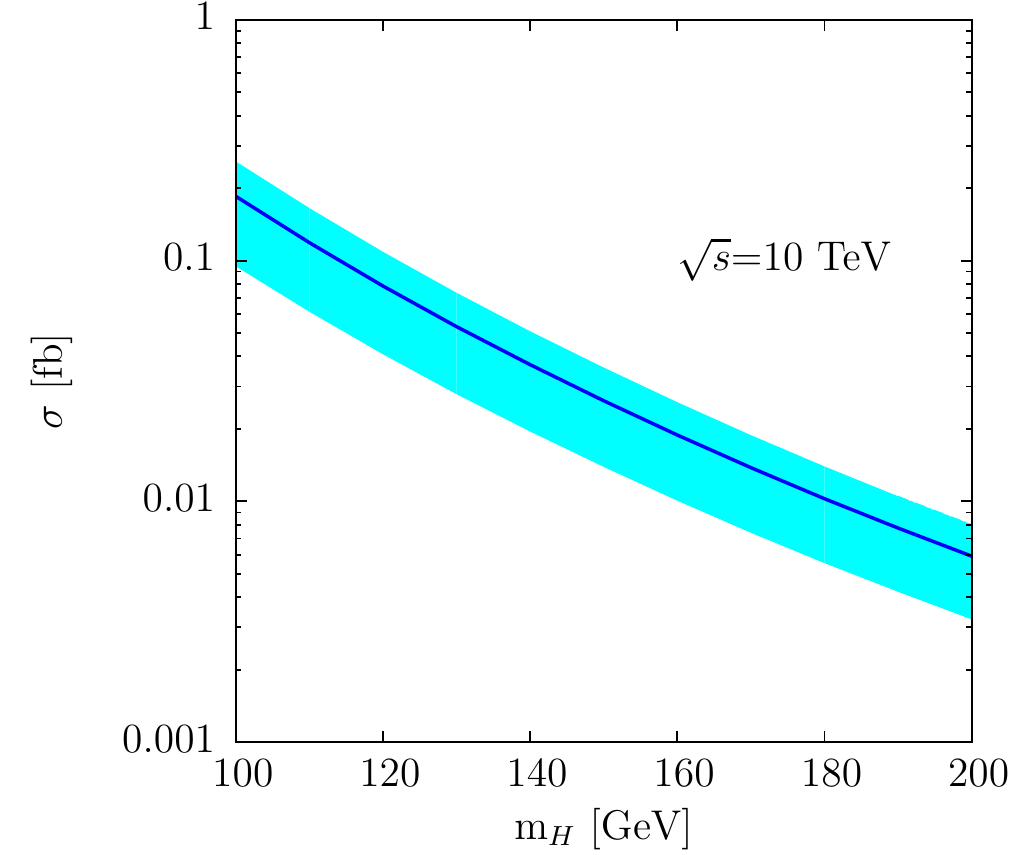} \includegraphics[height=4.3cm]{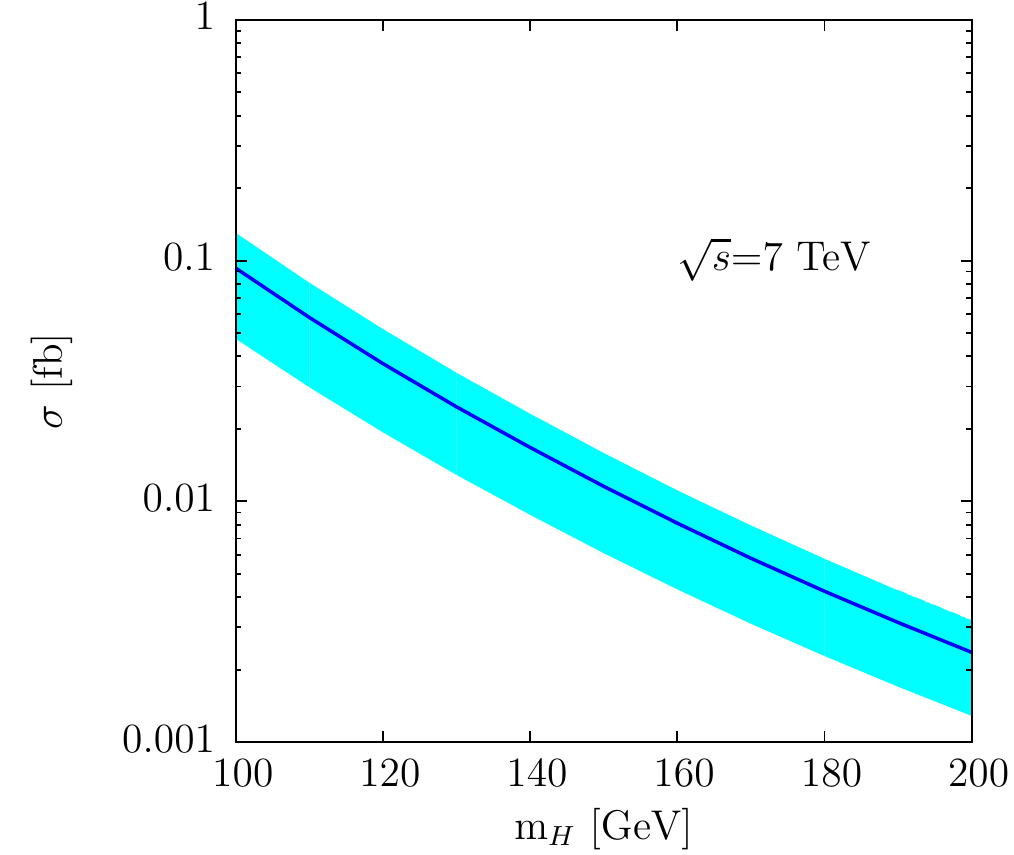} \\
\includegraphics[height=4.3cm]{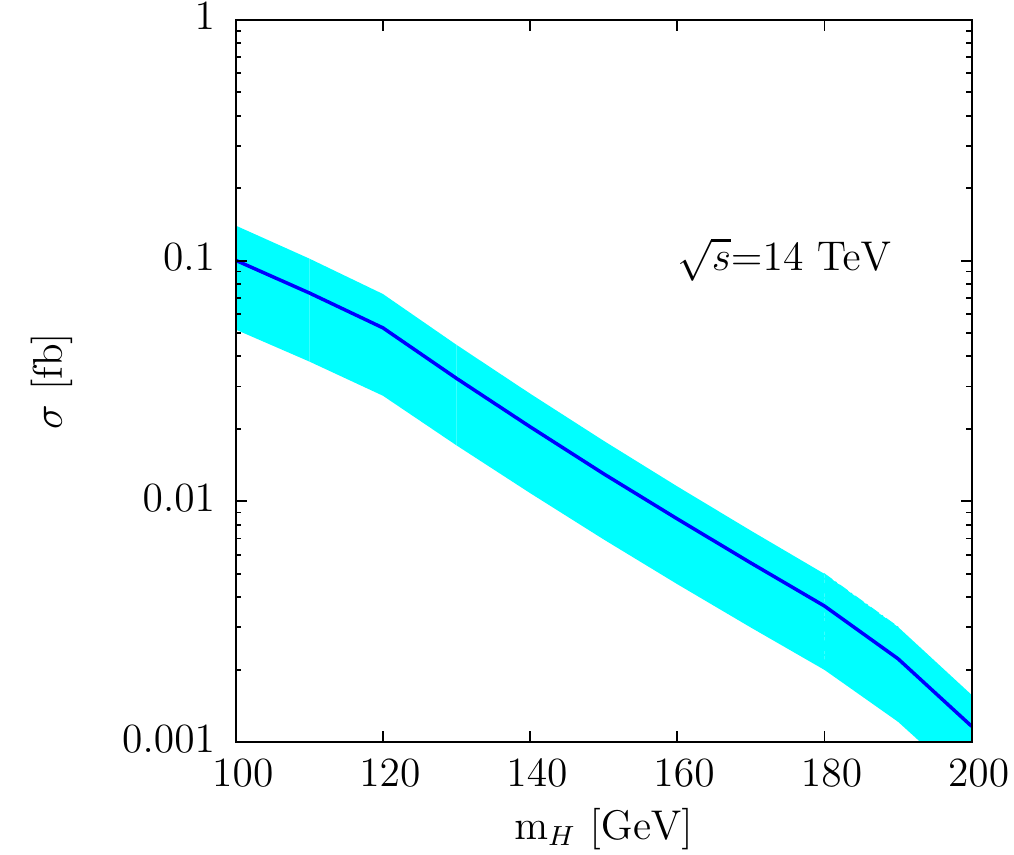} \includegraphics[height=4.3cm]{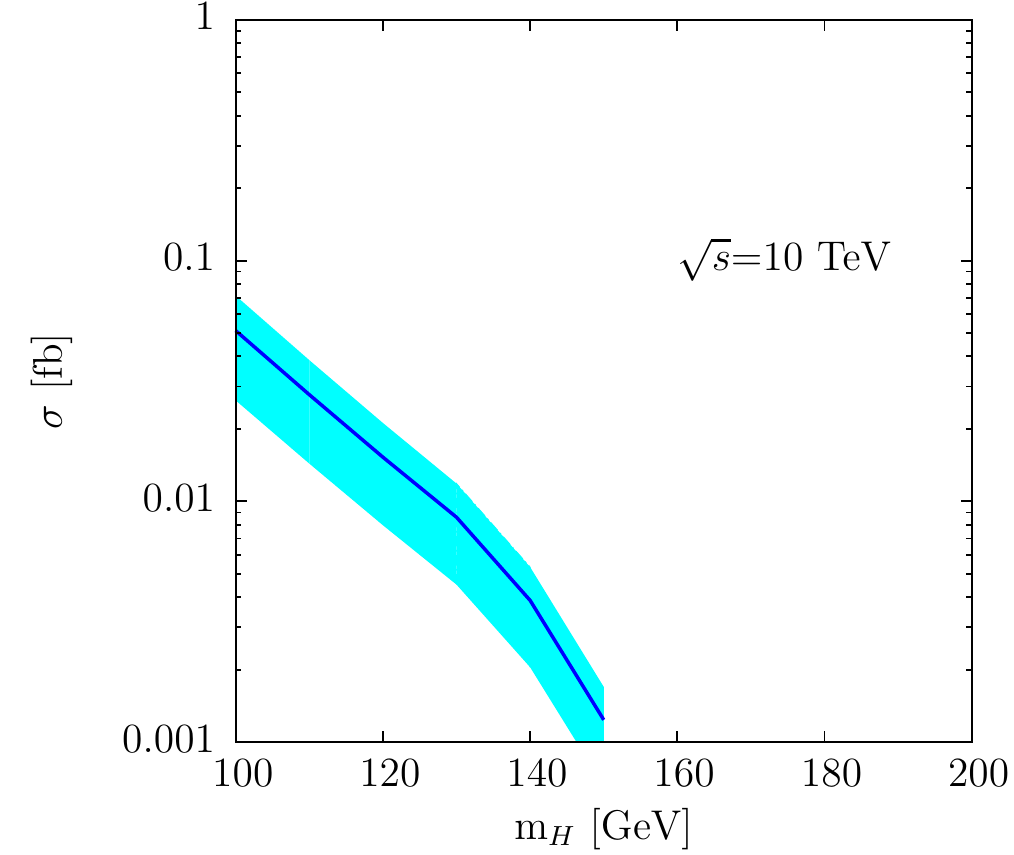} \includegraphics[height=4.3cm]{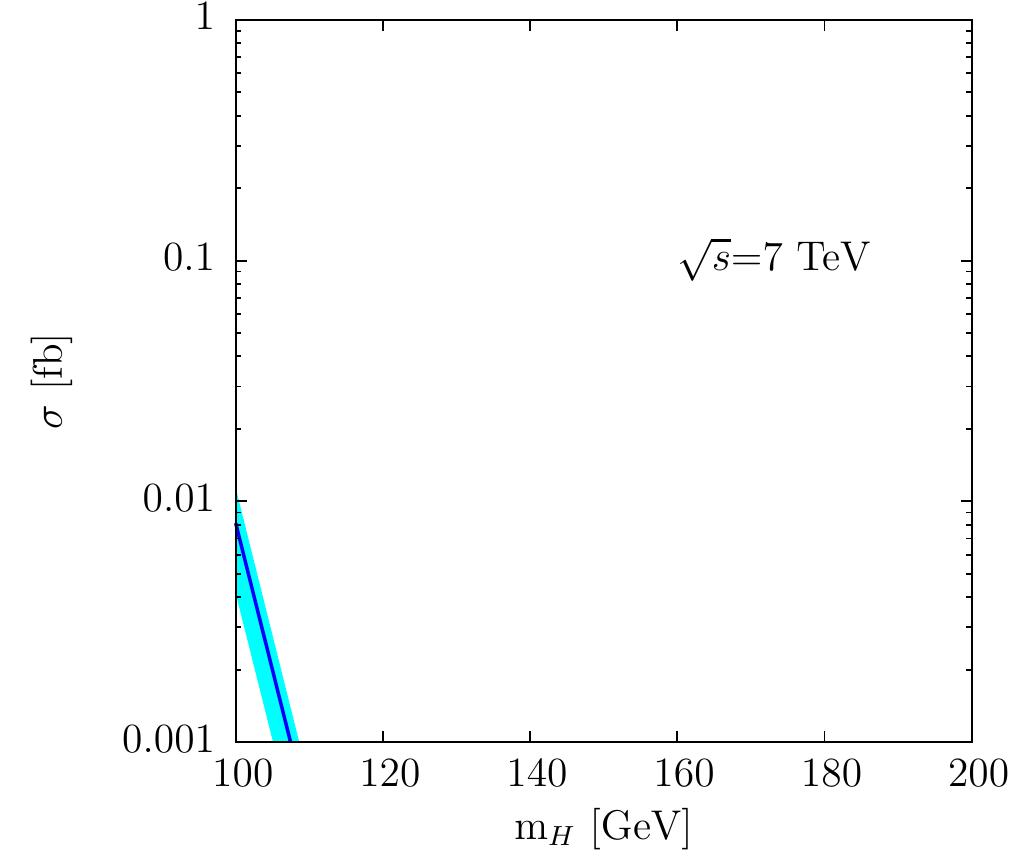}\\
\caption{Higgs-boson exclusive cross section at 14~TeV (left), 10 TeV (center) and 7~TeV (right). The first row gives the results without cuts. The second row corresponds to the cuts of the third column of Table \ref{table_cuts} and the third row to those of the fourth column}\label{fig_higgslhc}
\end{center}
\end{figure}
At the TeVatron ($\sqrt s$=1960~GeV) for the experimental cuts of the first column of Table \ref{table_cuts}, the Higgs boson exclusive cross section is tiny, of the order of 2~attobarn, for $m_H$=120~GeV and the parametrisation of the central curve. That makes the Higgs boson impossible to observe at the TeVatron, at least if it is produced in an exclusive reaction. 
\begin{table}[h]
\begin{center}
\begin{tabular}{cccc}
\hline  
 Cuts & TeVatron\cite{Aaltonen:2007hs}  & LHC\cite{Albrow:2005ig} & LHC\cite{Albrow:2008pn} \\
\hline
 $|\alpha_3|$     & [0.03,0.08]     & [0.002,0.2]   &  [0.005,0.018]  \\
 $\beta_1$        &  -              & [0.002,0.2]   &  [0.004,0.014]  \\
 $|y_X|$          & [-2.5,2.5]      & $<$ 1          &  $<$ 0.06       \\
 Gap size         & $>$ 3.6         &  -             &  -              \\
 M$_X$            &  -              &  $>$ 50 GeV    &  [80GeV,160GeV] \\
\hline
\end{tabular}
\caption{Experimental cuts for the TeVatron and the LHC.}\label{table_cuts}
\end{center}
\end{table}
For the LHC, we present in Fig.~\ref{fig_higgslhc} three sets of curves at 7, 10 and 14~TeV. The first row corresponds to the cross section without experimental cuts, the second for cuts resulting from the use of the forward detectors FP420+RP220~\cite{Albrow:2005ig}, given in Table~\ref{table_cuts}, and the third for a more restrictive set of cuts~\cite{Albrow:2008pn} used in Monte-Carlo studies, also given in Table~\ref{table_cuts}. The FP420 detector would be located in a small space left on the beam line of the LHC and be able to collect protons flying at 10 to 5~mm from the beam. With the addition of RP220 it should be possible to measure precisely the momentum of protons that have lost a fraction 0.002$<\xi<$0.2 of their initial momentum with an acceptance close to 100$\%$. This measurement would determine the mass of the centrally produced system and allow a clear identification of the exclusive events by associating the information from both central and forward detectors. 

In the left column that gives the cross section without any experimental cuts, one can see that the uncertainty was reduced by a factor~5 compared to Fig.~\ref{fig_maxhiggs} by using the CDF data. The effect of the cuts is important and cannot be neglected: the application 
of the FP420+RP220 cuts reduces the cross section by a factor three. The rapid fall down 
of the cross section in the bottom row is also purely due to the effect of cuts. If one 
considers only the use of FP420, there is a cut-off on the longitudinal momenta lost by 
the proton,

\begin{displaymath}
0.002 < |\alpha_3|,\beta_1 < 0.02.
\end{displaymath}
With $|\alpha_3|$ roughly equal to $m_H^2/s\beta_1$ if one considers a Higgs boson with 
a mass of 200~GeV in a process with $\sqrt{s}=10$~TeV, the only possible values for~$|
\alpha_3|$ stand between 0.02 and 0.2, and are outside the range of the forward 
detector. For collision of protons below 14~TeV, the use of the roman pots of RP220 in 
addition with the very forward FP420 is needed to increase the acceptance at large 
masses. This effect disappears at higher energies.

\par
\section{Conclusion}
In a previous paper~\cite{CDHI1}, we have proposed a new model for quasi-elastic 
production that is similar to the KMR model, but differs from it in the implementation 
of the different parts of the calculation: we use an impact factor with the correct IR 
properties, exact transverse kinematics and include a parametrisation of the 
nonperturbative region. Our model was used here to obtain the cross section of exclusive 
Higgs-boson production at the LHC and the corresponding uncertainties.
The vertex corrections,~{\it i.e} the Sudakov form factor, was one of the main sources 
of uncertainty in the dijet case but in the Higgs-boson production case, a 
recent~NLO~calculation sets its structure and scales. The other pieces of the model, the 
parametrisation of the unintegrated gluon density, the gap survival probability and the 
constant term in the Sudakov form factor were chosen to fit the CDF dijet data. Using 
these results, the Higgs-boson quasi-elastic cross section is found to be 
between~0.07 and 0.2~femtobarns for a Higgs boson of mass~120~GeV and~$\sqrt{s}=$~10~TeV 
considering the kinematical cuts of~RP220 and~FP420. This is a factor 2 lower than the 
KMR estimate~\cite{KMR}.

Finally, we want to stress that the theory of exclusive production is not under
full theoretical control. Indeed, exclusive production does not factorize, and wanders 
appreciably in the non-perturbative region. Hence, many of the assumptions made here are 
rather conservative as they derive from perturbative QCD. The constraint that leads to a 
reasonable agreement between models is the reproduction of the CDF dijet data.

\section{Acknowledgments}
The authors would like to thank I.~P~Ivanov for discussions and corrections. We also acknowledge discussions with M. Albrow, K.~A.~Goulianos, C.~Royon, and  thank J. Forshaw who clarified our ideas about the Sudakov form factor.

\end{document}